\magnification=\magstep1
\advance\vsize by1truecm
\voffset=-0.2truecm
\def\abstract#1{\vskip26truemm\noindent{\abstracttitlefont Abstract.}
{\abstractfont#1}\vskip16truemm}
\font\title=cmbx10 scaled\magstep1
\font\abstractfont=cmr9
\font\abstracttitlefont=cmti9
 \newcount\numref
\def\ref{\advance\numref by1${}^{\the\numref}$}
\overfullrule=0pt
\par\noindent
\centerline{{\title Un nouveau regard sur le monde
physique$^\ast$}}
\bigskip\bigskip\noindent
\centerline{Constantin Piron}
\bigskip\bigskip
\centerline{D\'epartement de Physique Th\'eorique}
\centerline{Universit\'e de Gen\`eve}
\centerline{CH-1211 Gen\`eve 4}
\par\noindent
\footnote{ }{$^\ast\,$Colloque de physique fait \`a Gen\`eve le 16 d\'ecembre 1996}
\abstract{Dans ce colloque je discute la notion d'espace selon Descartes et
Leibniz et j'y oppose le point de vue de Samuel Clarke et Isaac Newton
comme solution pour comprendre la r\'ealit\'e du monde physique }
\vskip2truecm
Au d\'ebut de ce si\`ecle la philosophie des sciences a \'et\'e secou\'ee
par deux r\'evolutions, les quanta et la relativit\'e. Je ne vais pas vous
d\'ecrire ici ces r\'evolutions, vous les connaissez\ref.
Mais paradoxalement pour la grande majorit\'ee des physiciens
l'image qu'ils se font du monde r\'eel est rest\'ee pratiquement
la m\^eme, celle d'il y a deux si\`ecles, celle des philosophes
Descartes et Leibniz. C'est pourquoi je vais tout d'abord vous r\'esumer tr\`es
bri\`evement le point de vue de ces deux philosophes, tel que je l'imagine \`a la
lecture de leurs textes. En fait, de Descartes, il ne faut retenir
que son d\'esir de rigueur math\'ematique et sa
conception m\'ecaniste de l'Univers, il faut oublier ses innombrables
tourbillons et surtout sa d\'efinition de l'essence des corps comme pure
extension sans vide d'aucune sorte. Leibniz \'etait moins dogmatique$\,$pour
lui l'espace n'est rien en soi ou
tout au plus un tissu de relations entre des objets coexistant. Comme
Max Jammer nous le rappelle dans son livre ``Concepts of Space''\ref ce point
de vue a d\'eja \'et\'e d\'efendu au 11$^{\hbox{i\`eme}}$ si\`ecle par le
philosophe musulman Al-Ghazale. Selon Aristote disait-il, le lieu
pr\'esuppose l'existence de corps, tout comme le temps, qui est le nombre du
mouvement. Or le dogme coranique affirme la cr\'eation divine des corps, le
temps et l'espace n'ont aucun sens "avant'' et ne sont donc que pures
relations entre les objets cr\'e\'es. Bien avant Galil\'ee,
Nicholas de Cusa
en 1440 avait d\'eja soustenu dans son trait\'e ``De docta ignorantia''\ref .
que tous les points de vue que nous prenons pour \'etudier l'univers
sont \'equivalents car il n'y a pas, disait-il, de centre du monde et donc
ni lieu ni temps absolu. Mais apr\`es
la d\'ecouverte par Newton des lois du mouvement et des lois de la
gravitation la situation s'est compliqu\'ee singuli\`erement. Il est
vrai que la fameuse exp\'erience du seau faite par Newton (son observation
de la concavit\'e de la surface de l'eau en rotation), n'a jamais
\'et\'e  consid\'er\'ee comme une preuve irr\'efutable
d'un mouvement absolu et en particulier d'un mouvement absolu de rotation.
Sur ce point, Leibniz est tout aussi
affirmatif, dans sa derni\`ere lettre \`a Huygens\ref  il \'ecrit$\,$:
``Je tiens donc que toutes les hypotheses sont
equivalentes et lorsque j'assigne certains mouuements \`a certains corps, je
n'en ay ny puis avoir d'autre raison que la simplicit\'e de
l'Hypothese croyant qu'on peut tenir la plus simple (tout consider\'e)
pour la v\'eritable''. Nos modernes, Einstein en t\^ete, ont
 adopt\'e compl\`etement cette th\`ese et ainsi il sont
accul\'es \`a devoir \'evoquer avec Mach l'effet lointain des \'etoiles fixes
pour pouvoir expliquer l'exp\'erience du seau de Newton.

D'autre part il faut rappeler que la loi de gravitation de Newton, son action
\`a distance, n'a
jamais \'et\'e acceptee (m\^eme par Newton), comme une loi
de nature fondamentale mais comme une loi devant pouvoir
s'expliquer m\'ecaniquement.
Dans l'Encyclop{\oe}dia Britannica\ref , \`a l'article ``Atoms'', Maxwell
\'ecrit que la meilleure explication connue est celle propos\'ee
par George-Louis Le Sage dans son trait\'e ``Physique m\'ecanique'' publi\'e
 \`a Gen\`eve en 1818\ref . Venant de l'ext\'erieur du monde et venant de
toutes les directions, un flot de tr\`es nombreux et tr\`es petits corpuscules
pousse tous les corps les uns contre les autres.
La loi de force qui en r\'esulte se trouve `au miracle' \^etre celle
propos\'ee par  Newton. Cette th\'eorie, dite des corpuscules ultramondains,
est revenue \`a la
mode vers 1870, a \'et\'e alors  abondamment  discut\'ee par les physiciens,
 m\^eme par des tout grands comme Sir William Thomson
(le futur Lord Kelvin) et Henri Poincar\'e. Cette th\'eorie, fut finalement
abandonn\'ee \`a cause de toutes les difficult\'es qu'elle
entra\^{\i}nait\ref . Mais elle ressemble curieusement \`a
la th\'eorie du graviton, l'hypoth\'etique particule associ\'ee
\`a la gravitation, qui est encore aujourd'hui la th\'eorie \`a la mode.

Dans un autre domaine, mais toujours dans la m\^eme ligne philosophique
l'\'ether luminifer a \'et\'e invent\'e pour combler le rien du vide et
pouvoir ainsi porter les ondes lumineuses de Maxwell.
Apr\`es les travaux  d'Einstein les physiciens ont appris \`a
se passer de l' existence d'un \'ether mais ils en ont gard\'e l'existence
d'un champ, une notion due \`a Faraday. N\'eanmoins, ils sont encore et
toujours accroch\'es \`a la m\^eme vision philosophique m\'ecaniste,
des petites particules en mouvement s'entrechoquant dans un ``rien''.
C'est ainsi que dans
presque tous les livres sur l'\'electromagn\'e\-tisme on peut lire, encore
aujourd'hui, l'argument suivant$\,$: Dans le vide il n'y a que le champ,
le champ $\vec{D}$ produit par une charge ne peut donc pas \^etre diff\'erent
du champ $\vec{E}$ qui agit sur l'autre charge. Cette identification brutale
de $\vec{E}$ et $\vec{D}$ et en cons\'equence de $\vec{H}$ et $\vec{B}$
conduit \`a bien des paradoxes. Par exemple la densit\'e volumique de
quantit\'e de mouvement du champ devient alors indiscernable de
la densit\'e de surface de son flux d'\'energie.

Si abandonnant les vues de Descartes et Leibniz on adopte celles de Newton
et Clarke et si on accepte l'existence en soi du vide et
l'existence en soi du temps, toutes ces difficult\'es disparaissent
d'elles-m\^emes. Aussi lais\-sez-moi vous pr\'esenter cette autre
philosophie du monde physique.

Comme Samuel Clarke l'affirme dans sa 4$^{\hbox{i\`eme}}$ r\'eponse \`a
Leibniz\ref$\,$: ``L'espace vide n'est pas un attribut sans sujet mais
un espace sans corps''. En termes modernes, l'espace vide n'est pas rien
mais un espace sans  quark ni lepton.
Dans le Scholium du chapitre Definitions au tout d\'ebut des Principia,
Isaac Newton\ref est tout aussi clair, apr\`es avoir mis en garde le
lecteur sur les erreurs de l'homme commun et avoir soigneusement
distingu\'e entre absolu et relatif, vrai et apparent, math\'ematique et
sens commun, il \'ecrit$\,$:
\medskip\noindent
{\parindent=40pt
\item{I} Le temps absolu vrai et math\'ematique de lui-m\^eme et de
par sa nature propre, coule uniform\'ement sans relation \`a rien d'ext\'erieur,
\item{II} L'espace absolu [vrai et math\'ematique] de part sa nature propre,
sans relation \`a rien d'ext\'erieur reste toujours similaire et immuable,
\par\noindent}%
\medskip\noindent

A mon sens ce Scholium doit \^etre replac\'e dans son contexte Platonicien.
Quand Newton \'ecrit ``absolu, vrai et math\'ematique'', il pense ce
que nous exprimons aujourd'hui par  ``physique et r\'eel'' et c'est ce qu'il
appelle le temps relatif et l'espace relatif qui pris ensemble est l' \^etre
math\'ematique. Car ${\bf R}^4\,$ est en effet la collection de tous les
 quadruples de nombres (les coordonn\'ees) construits \`a partir du choix d'un
rep\`ere au repos donn\'e. Comme l'affirme Einstein dans sa relativit\'e
g\'en\'erale ce choix est tr\`es arbitraire, cela peut-\^etre  le choix d'un
rep\`ere en chute libre choisi pour des raisons de
commodit\'e (par exemple la Terre ou le centre du syst\`eme
solaire)\ref. Mais c'est l\`a aller trop loin. En fait Einstein a
tort, car la lente et partiellement al\'eatoire d\'ecroissance de la vitesse
de rotation de la terre, mise en \'evidence seulement apr\`es les ann\'ees
60, a
d\'etruit irr\'evocablement l' argumentation de Mach en
fournissant enfin une preuve objective \`a la th\`ese de Galil\'ee
contre celle de Ptol\'em\'ee. Malgr\'e la forme de la dynamique
c'est bien la Terre elle-m\^eme qui tourne, car les \'etoiles lointaines ne
peuvent raisonnablement pas s'\^etre toutes donner le mot pour ralentir chacune
en son temps de mani\`ere \`a cr\'eer  l'illusion. La c\'el\`ebre
th\`ese de Minkowski qui explique la relativit\'e par la
g\'eom\`etrie d'un Univers absolu est
\'egalement en conflit avec l'exp\'erience, car l'\'etude
pr\'ecise du mouvement global des satellites a montr\'e qu'il n'y
avait aucune dispersion dans la coordonn\'ee temps de ces objets,
ceux-ci restent
confin\'es dans un espace ${\bf R}^3\,$ le sous-espace d\'ecrit par les trois
premi\`eres composantes du ${\bf R}^4\,$ consid\'er\'e\ref . Ce fait
d'exp\'erience ne peut s'expliquer que si l'univers ${\bf R}^4\,$ est non pas
d\'efini une fois
pour toutes, mais chaque fois reconstruit sur la base du rep\`ere au repos
choisi. Il est en effet toujours possible dans deux ${\bf R}^4\,$
li\'es par une transformation de Lorentz de d\'efinir dans chacun d'eux les
deux familles de sous-espaces qui vont repr\'esenter au cours du temps notre
 espace physique\ref.

Si donc l'existence du vide, si clairement impos\'ee par l'exp\'erience, est
n\'eanmoins ni\'ee par beaucoup, c'est qu'il existe
un fort pr\'ejug\'e contre ce concept parmi les philosophes et les physiciens
contemporains. C'est Newton lui-m\^eme qui a cr\'e\'e la confusion en comparant
l'espace \`a Dieu. Il a \'et\'e si loin dans son id\'ee qu'il a \'et\'e
conduit \`a \'ecrire\ref$\,$: ``[L'espace \'etendu] est un effet, une
\'emanation de l'\^etre existant primordial, car quand une chose quelconque
est l\`a, l'espace est l\`a''. En r\'esum\'e, d'apr\`es
Newton, l'espace existe car Dieu existe. C'est pourquoi les philosophes
du si\`ecle des lumi\`eres en ont imm\'ediatement conclu que l'espace
ne peut pas exister car Dieu n'existe pas\ref . Mais laissons l\`a cette
dispute philosophico-th\'eologique st\'erile et consid\'erons une toute autre
objection. Comment quelque chose qui serait juste ``rien'', une absence de
tout, pourrait avoir des propri\'et\'es tangibles? Car si ici, dans le vide,
nous pla\c cons un instrument nous n'avons plus le vide. Ce raisonnement
semble montrer l'impossibilit\'e logique d'appr\'ehender le vide en
lui-m\^eme et c'est l\`a apparemment une tr\`es s\'erieuse objection.

Apr\`es la d\'ecouverte des quanta et de la m\'ecanique quantique, les
physiciens ont \'et\'e forc\'es de discuter plus en pronfondeur ce que
signifie vraiment faire une mesure et tester une th\'eorie. Apr\`es bien des
discussions et gr\^ace aux nombreuses exp\'eriences r\'ealis\'ees, un point
de vue r\'eellement nouveau s'est d\'egag\'e. Ainsi
Diederik Aerts, dans sa th\`ese, a pu donn\'e un sens pr\'ecis au concept
de propri\'et\'e physique\ref . Il d\'efinit tout d'abord la notion de
projet exp\'erimental. C'est un projet pr\'ecis, aussi pr\'ecis que Bohr le
d\'esirait, d'une exp\'erience qu'on aurait parfaitement la possibilit\'e
de mettre \`a ex\'ecution et dont un des r\'esultats possibles a priori,
le r\'esultat dit positif, a \'et\'e choisi \`a l'avance. Aerts affirme
alors que le syst\`eme en lui m\^eme poss\`ede une propri\'et\'e
actuelle, un \'el\'ement de r\'ealit\'e aurait dit Einstein, si dans
l'\'eventualit\'e de l'ex\'ecution de ce projet, le r\'esultat positif
est par avance certain d'\^etre obtenu. Ce concept r\'esout la difficult\'e
en permettant de parler objectivement des propri\'et\'es
du vide. Ainsi par exemple, affirmer qu'ici, dans cette r\'egion,
le vide est Euclidien, c'est affirmer une propri\'et\'e actuelle du vide,
car si on construisait ici un triangle form\'e de trois solides r\`egles
rectilignes, tr\`es certainement, la somme des trois angles ainsi obtenus
serait \'egale \`a 180$^\circ\,$. C'est le vide lui-m\^eme en l'absence des
trois r\`egles qui poss\`ede cette propri\'et\'e. Ceci me rappelle une
conversation que j'ai eu \`a Copenhague avec Apostel durant une conf\'erence.
Il me disait dans l'\'enonc\'e de votre d\'efinition c'est le mot ``si'' qui
est important et qui joue le r\^ole clef, vous dites en effet$\,$:
Si vous ex\'ecutez cette exp\'erience vous obtiendrez ce r\'esultat. Les
raisonnements des physiciens, ceux qui conduisent \`a des paradoxes, sont
le plus souvent de faux raisonnements. Pour raisonner correctement en
physique il ne suffit pas d'appliquer na\"{\i}vement la logique \`a deux
valeurs mais il faut aussi tenir compte du fait qu'il y a une
troisi\`eme possibilit\'e, ne pas faire l'exp\'erience, et qu'ainsi le
r\'esultat cherch\'e est par nature hypoth\'etique.

Ayant ainsi accept\'e l'existence objective du vide en soi nous pouvons en
d\'efinir les propri\'et\'es et m\^eme les propri\'et\'es de nature non
purement g\'eom\'etriques. En fait ce que nous avons appel\'e un champ n'est
rien d'autre qu'une propri\'et\'e du vide en soi. Par exemple, ici, le champ
\'electrique $\vec{E}$ est un \'el\'ement de r\'ealit\'e, car si on
pla\c cait ici une charge \'electrique elle serait certainement
acc\'el\'er\'ee de mani\`ere pr\'ecise, en intensit\'e et en direction.
Comme nous l'avons d\'eja dit,  ce champ est le champ en l'absence de la
charge. Si on r\'ealisait cette exp\'erience le champ $\vec{E}$ serait
d\'etruit, en fait il ne serait m\^eme plus d\'efini. C'est l\`a la
mani\`ere correcte d'interpr\'eter les \'equations propos\'ees par Maxwell.
 Ainsi une charge \'electrique, par son champ
$\vec{D}$, va agir sur le vide qui va r\'eagir et produire le champ $\vec{E}$.
La difficult\'e mentionn\'ee n'existe pas les deux champs ne sont pas du tout
de m\^eme nature.

Mais tout ceci n'est pas suffisant pour pouvoir comprendre le monde
physique, le physicien doit encore changer sa conception sur la nature en soi
des particules. Cela est in\'evitable apr\`es la d\'ecouverte des
ph\'enom\`enes quantiques. L'\'electron n'est pas ce que la conception de
D\'emocrite pourrait sugg\'erer$\,$: un atome de substance \'electrique, un
point id\'eal charg\'e. S'il est vrai que l'\'electron est bien une entit\'e,
un objet sans partie n\'eanmoins il interagit non localement en se manifestant
comme un tout. De ce point de vue c'est plus une monade au sens de Leibniz
qu'un atome au sens de D\'emocrite. Ainsi l'atome d'hydrog\`ene au repos dans
l'\'etat fondamental doit \^etre con{\c c}u comme une boule ronde avec
le proton et l'\'electron, tous les deux \'egalement au repos et tous deux
non localis\'es tout autour du centre, le proton plus \`a l'int\'erieur et
l'\'electron plus \`a l'ext\'erieur\ref. Ce mod\`ele est tr\`es \'eloign\'e de
l'image plan\'etaire et n\'ecessairement plate qu'en a donn\'e Bohr dans sa
th\`ese. Dans ses discussions a propos de l'exp\'erience des deux fentes
de Young Heisenberg  pose \`a Bohr la bonne question$\,$: nous voulons savoir,
 dit-il, si la particule passe
par une des fentes, par l'autre ou par les deux ? C'est bien l\`a la bonne
question et apr\`es bien des exp\'eriences nous en connaissons maintenant la
r\'eponse, la particule ne passe, ni par une fente, ni par l'autre mais
par les deux \`a la fois\ref . N\'eanmoins le pr\'ejug\'e h\'erit\'e du
si\`ecle dernier est encore si fort que dans un journal s\'erieux comme
``Courrier CERN'' on pouvait lire r\'ecemment$\,$: ``Mais les \'electrons sont
des particules et de ce point de vue un \'electron doit traverser une
 fente ou l'autre, mais pas les deux\ref.

Cette nouvelle conception du monde physique o\`u une particule, comme
l'\'electron, se manifeste au cours du temps par des
propri\'et\'es particuli\`eres du vide lui-m\^eme\ref  est la bonne mani\`ere
d'interpr\'eter les ph\'enom\`enes physiques. De cette fa\c con
on peut comprendre, non seulement l'\'electromagn\'etisme mais aussi
la m\'ecanique quantique. Dans un article r\'ecent \'ecrit en l'honneur
de Hepp et Hunziker j'ai montr\'e comment en \'etudiant les \'equations
d'un champ du vide, le champ spinoriel, on pouvait d\'emontrer les
``r\`egles'' de la m\'ecanique quantique \`a partir de lois de conservations,
des lois du m\^eme type que celles qu'on d\'erive  des \'equations de
Maxwell pour l' \'electromagn\'etisme. Ainsi les r\`egles quantiques ne doivent
pas  n\'ecessairement \^etre impos\'ees de force de l'ext\'erieur, elles
d\'erivent naturellement des propri\'et\'es du champ du vide lui-m\^eme.
\parindent=0pt\bigskip
{\bf REFERENCES}
\bigskip
\numref=0
\def\refop#1{\hangafter=1\hangindent=30pt
\advance\numref by1{}\the\numref.\
#1.\par}
\def\refjl#1#2#3#4#5{\hangafter=1\hangindent=30pt
\advance\numref by1{}\the\numref.\
#1; ``#2'' {\it #3} {\bf #4} #5.\par}
\def\refbk#1#2#3{\hangafter=1\hangindent=30pt
\advance\numref by1{}\the\numref.\
#1; ``#2'' #3.\par}
\def\refpr#1#2#3#4#5{\hangafter=1\hangindent=30pt
\advance\numref by1{}\the\numref.\
#1; ``#2'' in #3 {\it #4} #5.\par}
\refpr{C. Piron}{Quanta and relativity: two failed revolutions}{}{Einstein
meets Magritte}{sous presse}
\refbk{M. Jammer 1969}{Concepts of Space}{Second Edition. p.50. Harvard
University Press, Cambridge}
\refbk{Nicholas of Cusa 1440}{De docta ignorantia}{Trans. G. Heron,
Routledge and Kegan Paul, London, 1954}
\refop{M. Jammer op. cit. p.123}
\refjl{J. C. Maxwell 1890}{Atom}{Encyclop{\oe}dia Britannica}{}{Ninth
Edition. Henry G. Allen, New York.}
\refbk{G.-L. Le Sage 1818}{Trait\'e de m\'ecanique physique}{Pr\'evost,
Gen\`eve.}
\refjl{S. Aronson 1964}{The gravitational theory of Georges-Louis Le
Sage}{Natural Philosopher}{3}{51-74}
\refbk{S. Clarke 1866 in}{Oeuvres philosophiques de Leibniz}{t.II, \S9, p.643.
Libraire Philosoph\-ique de Ladrange, Paris}
\refbk{Isaac Newton}{Principia Mathematica}{dans Opera qu\ae  exstant omnia
Vol. II, p.6. Friedrich Frommann Verlag (G\"unther Holzboog),Stuttgart-Bad
Cannstatt, 1964 }
\refbk{A. Einstein and L. Infeld}{The Evolution of Physics}{p. 224. University
Press, Cambridge, 1947 }
\refpr{C. Piron 1990}{Time, relativity and quantum theory}{A.
O. Barut (ed.)}{New  Frontiers in Quantum
Electrodynamics and Quantum Optics}{Plenum, New York}
\refjl{A. O. Barut, D. J. Moore and C. Piron 1994}{Spacetime models from the
electromagnetic field}{Helv. Phys. Acta}{67}{p.392-404}
\refpr{Cit\'e par J. E. McGuire}{Space, infinity, and indivisibility: Newton
on the creation of matter}{Z. Bechler (ed.)}{Contemporary Newtonian
Research}{D. Reidel, Dordrecht, 1982, p.160}
\refbk{Pour plus de d\'etails voir Voltaire}{El\'ements de la philosophie de
Newton mise \`a la port\'ee de tout le monde}{Chapter 2 ``De l'espace et
de la dur\'ee comme propri\'et\'es de Dieu''. Etienne Ledet,
Amsterdam, 1738}
\refbk{D. Aerts 1981}{The One and the Many}{Vrije Universiteit Brussel}
\refbk{C. Piron 1998}{M\'ecanique quantique bases et applications}{Presses
polytechniques et universitaires romandes (Lausanne). \S5.7.1}
\refop{Ibid \S1.5.23}
\refjl{G. Fraser r\'ed.1996}{M\'ecanique quantique. Le r\'eel rejoint le
virtuel}{Courrier Cern }{36{\rm:2}}{p.27}
\refjl{C. Piron 1996}{Quantum theory without quantification}{Helv. Phys.
Acta}{69}{694-701}

\bye